\documentclass[prl,aps,twocolumn]{revtex4}   
\usepackage{latexsym}
\usepackage{amsmath}
\usepackage{psfrag}
\usepackage{epsfig}
\newcommand{\be}[1]{\begin{align}\label{#1}}
\newcommand{\ee}{\end{align}}     
\newcommand{\bea}{\begin{eqnarray}}
\newcommand{\eea}{\end{eqnarray}} 
\newcommand{\eq}[1]{Eq.~(\ref{#1})}

\newcommand{\fig}[1]{Fig.~\ref{#1}}
\begin{document}

\tighten   
%
\title{ \large\bf Electron release of  rare gas atom clusters under an 
intense laser pulse}
\author{Christian Siedschlag and Jan M. Rost} 
\affiliation{
 Max-Planck-Institute for the Physics of Complex Systems, 
N\"othnitzer Str. 38, D-01187 Dresden, Germany}
\date{\today}
\begin{abstract}
Calculating the energy absorption of atomic clusters as a function of
the laser pulse length $T$ we find a maximum for a critical $T^*$.  We
show that $T^*$ can be linked to an optimal cluster radius $R^*$.  The
existence of this radius can be attributed to the enhanced ionization
mechanism originally discovered for diatomic molecules.  Our findings
indicate that enhanced ionization should be operative for a wide class
of rare gas clusters.  From a simple Coulomb explosion ansatz, we
derive an analytical expression relating the maximum energy release to
a suitably scaled expansion time which can be expressed with the pulse
length $T^*$.
\end{abstract}
\pacs{PACS numbers:  36.40.-c, 33.80.-b, 42.50.Hz}
\maketitle
After a basic understanding of  the mechanisms governing atoms and 
molecules subjected to an intense laser pulse \cite{Joach,McCaPost99},  
analogous studies on clusters pioneered by Rhodes \cite{mcp1} and Ditmire 
\cite{ditmire} have appeared over the last years with a recent spectacular 
culmination  in the demonstration of deuterium fusion in clusters 
\cite{Dital97}. Most of these studies 
do focus on the  situation after the laser pulse, namely on the abundance 
and kinetic energy spectra of electrons and ions. Some discussion has 
been devoted to the question if the expansion of the cluster is driven 
by hydrodynamics or by a Coulomb explosion. Only very little attention 
has been paid to this type of dynamics in the time domain 
\cite{LaJo00,IsBl00}. This is even more surprising since the time scales 
involved show that the expansion of the nuclei occurs on the same time 
scale as the pulse lengths which can be chosen, namely some 10 to 
1000 fs, or roughly $10^{-3}$ atomic units (which we will use 
hereafter).  Apart from the nuclear motion and the pulse length $T$ 
energy absorption from a laser pulse and subsequent 
ionization and fragmentation of the cluster involve two
additional time scales,  the optical cycle $2\pi/\omega = 
0.055$a.u. for  the typically used Titan-Sapphire laser of 800 nm wavelength,
and the period of the bound electrons, which is of the order (hydrogen) 
of  1 a.u..  We will work with peak intensities between $10^{14} 
-10^{16}$ W/cm$^2$.\\
In the following we will demonstrate that the seemingly complicated 
process of energy absorption and fragmentation in the laser pulse can 
be split into three different phases, an 
`atomic '  phase I,  a `molecular' phase II, and a relaxation phase III.
Phase I lasts for a time $T_{0}$ after the pulse has begun and is 
characterized by boiling off electrons through multiphoton or  
tunneling ionization, hence we have termed it `atomic' phase. We 
define it to last until every second atom in the cluster has lost 
one electron, or equivalently until the probability of loosing an 
electron in an atom has reached  $p = 1/2$. This probability is 
calculated from a Krainov tunneling rate \cite{amm1} where, however, 
the instant electric field is formed by the laser and eventually 
already existing charged particles in the cluster.

Up to $T_{0}$ we may assume that the atoms/ions have not moved yet.
The second, molecular phase is characterized by Coulomb explosion 
of the cluster. During this phase, as we will show below, the cluster expands
to a critical radius $R^{*}$ which optimizes the energy absorption.
Phase III finally, until the end of the laser pulse and beyond, sees 
a relaxation of the system and the full fragmentation of the cluster 
proceeding.
The existence of these phases follows from a careful analysis of our numerical
results. The relevance of the phases is underlined by the time which is spent
under phase II. This time turns out to be instrumental for relating the
electron release quantitatively to the laser and cluster properties, as will
be shown below.

To simulate the process of energy absorption numerically we have
developed a quasiclassical model for small rare gas clusters.  The
nuclei are treated completely classically, with the initial
configuration defined by minimizing the pairwise Lennard-Jones
interactions \cite{LenJon}.  Electrons bound to an
atom or ion at position $\vec{R}$  are characterized by an effective binding energy
\begin{align}
E_b=E_b^{{\rm Atom}}+V_{{\rm total}}(\vec{R})
\end{align}
where the exact atomic binding energy $E_b^{{\rm Atom}}$ is shifted by
$V_{{\rm total}}$, the sum of the potentials from the laser
field and all other charged particles except the atom/ion the electron is
bound to. 
 Ionization from such a bound state is accomplished via
tunneling along the direction $\hat r$ of the instant force at position 
$\vec{R}$ , $\hat r = \vec{\nabla} \ V(\vec{R})/|\vec{\nabla} V(\vec{R})|$.  The time-dependent tunneling 
action along $\hat r$ reads
\begin{align}
I(t)=\int_{r_1}^{r_2} \sqrt{2(V_{{\rm total}}(r)+V_{{\rm Atom}}(r)-E_b)}\ dr
\end{align}
with the classical turning points 
$r_i$  determined by $V_{{\rm total}}(r)+V_{{\rm Atom}}(r)-E_b=0$. 
From $I(t)$ we get the tunneling probability $P(t)=\exp(-2 I(t))$ 
and finally the tunneling rate 
\begin{align}
w(t)=\frac{1}{T_{K}}P(t)
\end{align}
with the classical Kepler period $T_{K}$ of an orbit with binding
energy $E_b^{{\rm Atom}}$.  For each time step $dt$, a random number
$z$ is compared to the probability $w(t) \ dt$ for ionization during
this time step.  If $w \ dt>z$, the electron is born as a classical
particle and placed at the outer turning point $r_2$ obeying total
energy conservation.  From then on, this electron follows Newton's
equations, and the next bound electron can be ionized.  Hence, 
  strictly sequential ionization is enforced.

 The interaction between two particles with charge $Q_1$ and $Q_2$ and
 position vectors $\vec{r}_1$ and $\vec{r}_2$, respectively, is
 described with a smoothed Coulomb potential
\begin{align}
V_{Softcore}=\frac{Q_1 
Q_2}{\sqrt{(\vec{r}_1-\vec{r}_2)^2+a_1(Q_1)+a_2(Q_2)}}\,,
\end{align}
where the $a_i$ are charge-dependent soft-core parameters.  For
electrons we used $a(-1)=0.1$, while the ionic $a_i$ are chosen such
that the potential minima for each ion always coincide with the
quantum mechanical binding energy.  This choice prevents artificial
classical autoionization.

The model allows us to follow the full time-dependent evolution 
of the cluster with all interactions for a long time ($10^5$ a.u.) 
to investigate the influence of
the cluster expansion during the laser pulse on its energy 
absorption. Although it implies, e.g.\ for xenon clusters, to 
propagate up to 200 charged particles,  the computation can be 
handled with moderate resources due to the crucial simplification 
which arises from treating bound electronic motion not explicitly.

\begin{figure}
\centering
\psfrag{xtitle}[][][1]{T}
\psfrag{ytitle}[][][1]{atomic charge}
\epsfig{file=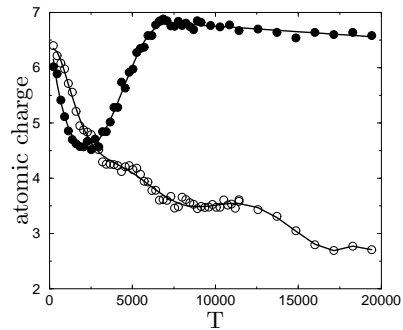,width=0.6\columnwidth}
\caption{Average atomic charge of Xe$_{16}$ ($\bullet$) and Xe ($\circ$) as a function of pulse
length. The lines are to guide the eye.}
\label{Xe}
\end{figure}

After the pulse is over the electron release from the cluster is a 
typical observable which changes as a function of the pulse length $T$
as shown in \fig{Xe} for a Xe$_{16}$ cluster in 
comparison with the corresponding electron release (i.e., ionization) of a Xe atom. 
The energy content of the 
laser pulse ${\cal E }= \int_{0}^{T} F(t)^2 dt$ has been kept constant which 
means that the peak intensity $F_{0}^2$ of the  pulse with
amplitude $F(t)=F_{0}\sin^2(\pi t/T)\cos\omega t$ decreases with increasing 
pulse length $T$ according to $F_{0}^2\propto 1/T$. As a reference for
this energy normalization we chose a pulse with 
$F_0=0.16$ a.u. and a pulse length of 20 optical cycles. One sees that the 
ionization of an atom increases towards shorter pulse lengths 
$T$ or equivalently, higher peak intensity. Indeed, 
atomic ionization depends on the peak  intensity  
$F_{0}^2$ rather than on the pulse length $T$ which is obvious if the
electron yield is dominated by sequential ionization 
 depending exponentially on $F_0$ via the Krainov rate \cite{amm1}, but
only linearly on the pulse length. The oscillations in the single atom 
case are due to the atomic shell structure.

\begin{figure}
\centering
\psfrag{xtitle}[][][1]{$R/R_0$}
\psfrag{ytitle}[][][1]{atomic charge charge}
\epsfig{file=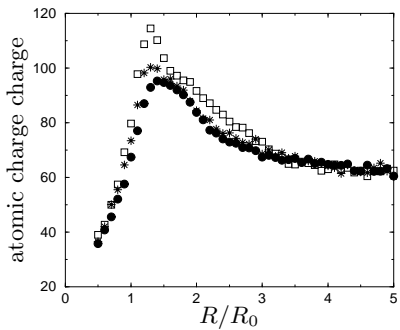,width=0.6\columnwidth}
\caption{Electron release 
of Xe$_{16}$ for fixed nuclei. Results are shown for 
for   a pulse length
of $T=55$ fs and the frequencies $\omega=0.055$ a.u.\ ($\bullet$),
$\omega=0.075$ a.u.\ ($\star$), and $\omega=0.11$ a.u.\ ($\Box$).
}
\label{statmax}
\end{figure}

For the cluster the situation is quite different: for short $T$ (high
intensity $F_0^2$) the electron release/atom follows that of an
isolated atom. In fact, it is even slightly lower than in the atomic
case, due to the much larger Coulomb field of a multiply ionized 
cluster which has not
significantly expanded. However, the electron release
increases again and reaches a
maximum for some optimum pulse length $T^*$ at considerably smaller
peak intensity.  Hence, the cluster \emph{expansion} plays an important
role for the energy absorption, in contrast to the 
atom for which this degree of freedom does not exist.  A comparison of
time scales shows, that indeed the Coulomb explosion of the ions in
the cluster happens on the same time scale as the pulse duration
($10^3-10^{4}$ a.u. or equivalently some $10 $ to $100 $ fs).  Hence,
the dependence of the electron release on $T$ points
to the spatial expansion of the cluster which may in turn exhibit a
maximum electron release for a certain cluster radius $R$.  We 
define $R$ in terms of the averaged distance between two ions in
the cluster,
\begin{align}
\label{R}
R(t) = \left(\frac{1}{N}\sum_{i=1}^N \min_{i \neq j} \{|\vec R_{i}-\vec
R_{j}|^2\}\right)^{1/2}\,.
\end{align}
First we assess the influence of the size of the cluster on the 
electron release under the  reference
pulse of 20 field cycles. The size of the cluster is varied 
preserving its shape by scaling the 
ionic positions $\vec R_{i}^{\lambda} = \lambda 
\vec R_{i}^{0}$ with a factor $\lambda$ compared to the equilibrium 
positions $\vec R_{i}^{0}$.
As can be seen from \fig{statmax},   a
critical value of the mean interionic distance, $R^{*}=\lambda^{*} R_0$ exists, where the ionization yield shows a maximum. 
The position of $R^{*}$ hardly changes upon 
variation of the laser frequency.
The ionization yield, however, increases with increasing frequency:
this is due to the smaller ponderomotive oscillation amplitude at higher
frequencies, which leads to increased interaction between quasi free 
electrons driven by the laser field and those  still well in reach of 
the cluster ions.

\begin{figure}
\centering
\psfrag{xtitle}[][][1]{$R/R_0$}
\psfrag{ytitle}[][][1]{atomic charge}
\epsfig{file=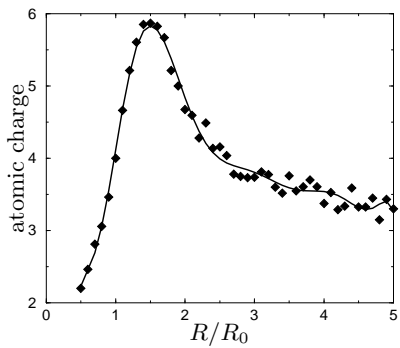,width=0.6\columnwidth}
\caption{Electron release of Xe$_{16}$ with fixed nuclei and circular polarization.}
\label{circular}
\end{figure}

The mechanism responsible for the existence of $R^{*}$ was
first identified in diatomic molecules under the name CREI or enhanced
ionization (ENIO) \cite{Seial95,ZuBa95} with the (linear) laser
polarization parallel to the molecular axis.  It might seem
astonishing that the cluster also exhibits ENIO although there is no
preferred axis which could align with the polarization axis.  This is
even more surprising since no enhancement was found for diatomic
molecules if the polarization is perpendicular to the molecular axis
or if the laser is circularly polarized.  However, a distinct feature
of ENIO is the insensitivity to changes in the laser frequency which
we also find in the cluster (\fig{statmax}).  This fact, together with
the relation of $T^*$ to the critical radius $R^*$ as presented
below provide sufficient evidence that intense laser field dynamics of
clusters is structured by ENIO as is the corresponding dynamics of
molecules.  For clusters, ENIO is even more general since there is no
restriction with respect to the polarization of the laser: The
direction of the axis for linear polarization does not matter and ENIO
also occurs for circular polarization as shown in \fig{circular}.

Our findings exemplified here for Xe$_{16}$ have been confirmed by 
extensive calculations for a number of clusters of 8 to 30 atoms for 
the elements Ne, Ar, Kr, and Xe. These calculations clearly demonstrate
 that ENIO plays an important role for
 small rare gas clusters under intense laser fields with
quantitative consequences as we will see next.

\begin{figure}
\centering
\psfrag{xtitle}[t][][1.1]{$t$}
\psfrag{ytitle}[][][1]{$R(t)$}
\psfrag{t1}[][][0.8]{I}
\psfrag{t2}[][][0.8]{II}
\psfrag{t3}[][][0.8]{III}
\psfrag{t4}[][][0.8]{$T_0$}
\psfrag{t5}[][][0.8]{$T$}
\psfrag{t6}[l][][0.8]{$R^*$}
\epsfig{file=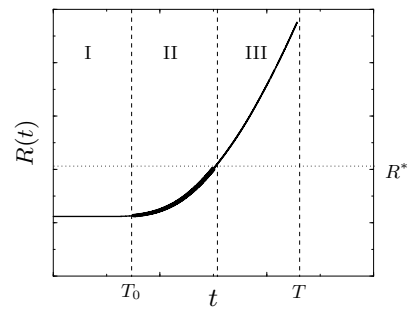,width=0.6\columnwidth}
\caption{Sketch of phases I, II and III during the pulse (see text)}
\label{sketch}
\end{figure}

One could think that the relation of $T^*$ and $R^*$ is directly given
by the Coulomb explosion mechanism.  The latter links indeed $R^*$ to
a certain time interval $\tau$, but not to the entire pulse length
$T^*$.  The reason lies in the existence of the different phases as
described in the beginning and schematically shown in \fig{sketch}. 

Only phase II drives the relevant Coulomb explosion,
namely the onset of the cluster expansion.  For the maximum electron
release this time interval ends if the critical radius $R^*$ is
reached at half the pulse length $T^*/2$ when the laser pulse
intensity has its maximum.  The interval begins, however, only at
$T_{0}$ when each atom in the cluster has on average a 50\%
probability of being ionized.  At this time the cluster still has its
equilibrium radius $R_{0}$.  Hence, we get $\tau = T^*/2-T^*_0$ as the
relevant time during which the  cluster expands from $R_{0}$ to $R^{*}$.

Having identified phase II and its time interval $\tau$ as the one 
which  controls the maximum number $Q^*$ of released electrons, we can 
use the dynamics of Coulomb explosion to derive a quantitative 
relation between $Q^*$ and the optimum pulse length $T^*$.
To this end we assume that the ionic motion in $\tau$ can be 
described by an averaged ionic charge which is proportional to the 
averaged final charge of each atom in the cluster, i.e. 
\begin{align}
\label{charge}
Q_{i}= \alpha Q^*/N,
\end{align}
where $N$ is the number of atoms in the cluster.
Furthermore we assume that in phase II the main kinetic energy goes into the 
expansion of the cluster  without changing its shape. Using the same 
parameterization as before, we write now for the time-dependent 
position $\vec{R}_{i}(t)$ of an atom or ion $\vec{R}_{i}(t) = 
\lambda(t) \vec{R}_{i}^{0}$, where $\vec{R}_{i}^{0}$ is the equilibrium position of 
the atom before expansion, i.e., at time $t = T_{0}$. The kinetic 
energy  $K = M/2\sum_i (d\vec{R}_{i}(t)/dt)^2$
reads then
\begin{align}
\label{kes}
K= \left( \frac{d\lambda}{dt} \right)^2 \frac{M}{2}\sum_{i}^{N}(\vec{R}_{i}^0)^2
\equiv \left( \frac{d\lambda}{dt} \right)^2 I_{0}\,.
\end{align}
$I_{0}$ has  form and units of a moment of inertia and represents the
influence of the shape of the cluster on its kinetic energy during 
the expansion.
The potential energy $V = \sum_{i>j}Q_{i}Q_{j}/|\vec{R}_{i}(t)-\vec{R}_j(t)|$ simplifies to 
\begin{align}
\label{POT}
V = \lambda^{-1}(\alpha Q^*/N)^2\sum_{i>j=1}^{N}|\vec{R}_{i}^0-\vec{R}_{j}^0|
\equiv \alpha^2 V_{0}/\lambda\,.
\end{align}
The differential equation for the expansion in terms of $\lambda(t)$
is obtained via the energy balance $K(t)+V(t) = E \equiv V(T_{0})$,
where at time $T_{0}$ before the expansion the kinetic energy of the
atoms is zero.  With the help of \eq{kes} and
\eq{POT} it can be written in the 
form 
\begin{align}
\label{diflam}
\frac{d\lambda}{dt}= \alpha [(1-\lambda^{-1})V_0/I_{0}]^{1/2}\,.
\end{align}
Eq.~\ref{diflam} can be solved analytically by separation of variables to yield
\begin{align}
\label{lam}
t (\lambda) -T_{0}= \left(\frac{K_{0}}{V_{0}\alpha^{2}}\right)^{1/2}f(\lambda)
\end{align}
where we have set $\lambda =1$ for $t=T_{0}$ and $f(\lambda):=\sqrt{\lambda(\lambda-1)}+\ln(\sqrt{\lambda-1}+\sqrt\lambda)$.  For the maximum
energy release the critical radius $R^* = \lambda^{*}R_{0}$
should be reached after time $t(\lambda_{\rm crit})= T^*/2$.  This is
the desired relation between the static ENIO mechanism at $R^*$ and
its dynamical effect during the cluster expansion in the time interval
$\tau = T^*/2-T_0^*$.

The proportionality
factor  in \eq{charge}  determines the fraction of the final charge by
which the expansion from $R_0$ to $R^*$ during phase II is effectively driven.
If phase II is indeed the crucial dynamical time span which 
universally controls 
the electron release we expect  $\alpha$ to be
the same for all types of clusters we consider, independent of the atomic 
element or cluster size. Under this assumption, we predict from \eq{lam} a linear
relation between the expansion time $\tau$ and
$(K_{0}/V_{0})^{1/2}f(\lambda)$.
\begin{figure}
\centering
\psfrag{xtitle}[t][][1]{$(K_{0}/V_{0})^{1/2}f(\lambda)$}
\psfrag{ytitle}[][][1]{$\tau$}
\epsfig{file=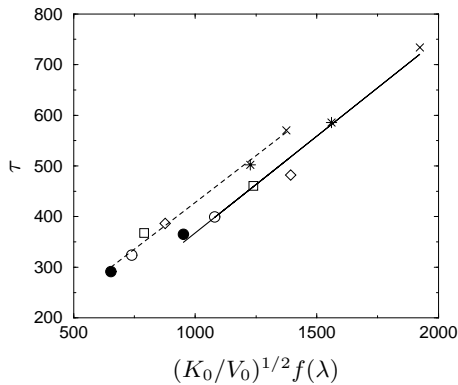,width=0.7\columnwidth}
\caption{Expansion time (numerical data) as a function of
$(K_{0}/V_{0})^{1/2}f(\lambda)$ and linear fits (see text). Two different energy normalizations
were used: $F_0=0.16$ a.u. (solid line) and $F_0=0.25$ a.u. (dashed line), both at a frequency of
$\omega=0.055$ a.u. and a pulse length of $T=55$ fs.   
$\bullet$: Ar$_{16}$, $\circ$: Ar$_{20}$, $\Box$: Ar$_{25}$, $\Diamond$: Ar$_{30}$,
$\star$: Kr$_{16}$ and $\times$: Xe$_{16}$}
\label{linear}
\end{figure}
In \fig{linear} we can see that this prediction is actually very good:
shown are the expansion times $\tau$ as a function of the
cluster-dependent values of $(K_0/V_0)^{1/2}f(\lambda)$ for different
clusters.  A linear fit to the data yields $\alpha=0.38$ and
$\alpha=0.37$ for energy normalized pulses at $F_0=0.16$ and $F_0=0.25$,
respectively.  The correlation coefficient is in both cases higher
than 0.99.  Hence, $\alpha$ is the same for different clusters, and it is
almost the same for different energy normalizations of the laser
pulse.

To summarize, we have shown that the enhanced
ionization mechanism is operative for small rare gas clusters over a wide
range of parameters. Moreover,
from a careful analysis of the Coulomb explosion process, 
we  conclude that energy
absorption and subsequent ionization of the cluster proceeds in a very similar
way for different clusters, 
irrespectively of the number and sort of atoms in the cluster.
It is only for large rare gas clusters, with $N\sim 10^3$
or more, that we expect a transition to a nanoplasma behavior,
as it has been found in hydrodynamical simulations of such systems
\cite{ditmire,zwei}. Where and how this transition happens will be the 
subject of further studies as well as the connection with enhanced 
energy absorption recently reported 
for small {\em metal} clusters \cite{meiwes}.

\bibliographystyle{unsort}

\end{document}